# On the long-term health care crisis. A possible eradication scenario.

Raul Isea[1,*], Er Wei Bai[2] and Karl E. Lonngren[2]

[1] Fundación IDEA, Hoyo de la Puerta, Baruta 1015, Venezuela.

[2] Department of Electrical and Computer Engineering, University of Iowa, Iowa City, IA 52242-1527.

(*) Corresponding author: Raúl Isea. Fundación IDEA, Hoyo de la Puerta, Baruta, Venezuela.

## Abstract

The purpose of the present essay is to suggest a possible model to describe the worldwide healthcare crisis, where diseases that have been considered to be eradicated or under our control are re-emerging today.

Poliomyelitis that once plagued humanity is still a pandemic disease again (Newsom 2005). In 1916 in the USA, there were 23,000 documented cases causing more 5,000 deaths (Trevelyan *et al.* 2005). Moreover, the poliomyelitis epidemic began to grow after World War II (Trevelyan *et al.* 2005), and for this reason, the 41st World Health Assembly meeting in Geneva in 1998 committed the World Health Organization (WHO) to commence a global campaign to eradicate poliomyelitis (Goodman 2000). The goal has not been achieved as there have been recent outbreaks in Congo (CDC 2011), Tajikistan (Macdonald & Hébert 2010) and Venezuela (Echezuría 1974), and a clear example shown in the Figure 1 illustrates a recent poliomyelitis outbreak in Pakistan.

Another example of an epidemic is malaria. It is estimated that one half of the worldwide population is at risk of being infected with malaria with two to three million people dying annually as a result of it. In 2007, the Pan American Health Organization (PAHO) reported that the disease had affected twenty-one countries throughout the Americas with approximately 880 million people of which 236 million live in endemic areas and 276 million live in areas at risk of transmission (Isea 2010). Currently, there is hope that an effective malaria vaccine called RTS, S/AS02 will prove effective based on studies performed in Mozambique in 2009 (Sacarlal 2009). The malaria parasite mutates to avoid the effects of this vaccine and has proven to be effective for only a limited duration of approximately 45 months (Sacarlal 2009).

Based on these two examples, we raise a simple question: <u>Is it possible to eradicate diseases worldwide or is there a cyclical nature to the increase and decrease of a pandemic</u>

disease? A possible answer to this question may be obtained from an analysis of a modification of the predator-prey equations as will be demonstrated below.

Let us assume that the dependent variable *well* represents the population of well people and the dependent variable *sick* represents the population of sick people that reside on the earth. At the present time, the overall population is increasing in time due to the increasing birthrate throughout the world. This increase in the well population will be represented with the coefficient $\alpha$. The increasing population of well people will encounter a number of sick people who could infect the well people and the infection rate is represented with the coefficient $\beta$. We would expect a certain proportion of the sick population will recover and become members of the well population and this is represented with the coefficient $\gamma$.

$$\frac{d(well)}{dt} = \alpha\,(well) - \beta\,(well)(sick) + \gamma\,(sick) \tag{1}$$

The population of sick people will decrease as the sickest of the sick pass away and are buried in the ground. This is described with the coefficient $\delta$. The burial of the sick population fertilizes the ground that will in turn enhance the food production for the well people and their population will be in turn affected and this is described with the coefficient $\varepsilon$.

$$\frac{d(sick)}{dt} = -\delta\,(sick) + \varepsilon\,(sick)(well) \tag{2}$$

Equations (1) and (2) have been recently examined with reference to the problem of global warming caused by the production of carbon dioxide (Lonngren & Bai 2008). In addition, there has been a detailed study of the mathematical behavior of these two equations in the regions surrounding the equilibrium points (Schaffer 2008).

Asymptotically, there are two long-term equilibrium points for these equations. The first one is the equilibrium point at the origin, ie., *well* = 0 and *sick* = 0.

The second equilibrium point is given by.

$$well = \frac{\delta}{\varepsilon} \quad \text{and} \quad sick = \frac{\alpha\left(\frac{\delta}{\varepsilon}\right)}{\beta\left(\frac{\delta}{\varepsilon}\right) - \gamma} \quad (3)$$

Equations (1) and (2) with the coefficient $\gamma = 0$ are just a reinterpretation of the well-known predator-prey model equations. These equations have been used to describe the temporal evolution of the population of rabbits (prey) and the population of foxes (predator) in a closed environment such as an island. As the population of the rabbits increases, they in turn become food for the foxes. As the food supply for the rabbits increases, the population of the foxes also increases resulting in an eventual decrease in the population of the rabbits. The decrease of the rabbit population will then cause a decrease in the population of the foxes and the scenario repeats itself. This process is also called the Lotka-Volterra model in honor of their independent description of these equations in 1925-26 (Berryman 1992).

The solution of the predator-prey equations (or Lotka-Volterra model) is shown in Figure 2 where we have arbitrarily chosen numerical values for the coefficients of equations (1) to (3) and thus emphasize the oscillatory behavior that may extend over thousands of years. The cases of polio in Pakistan exhibits such a tendency as shown in Figure 1. The numerical values that were used in the simulation are included in the figure caption. The major result of this calculation is to demonstrate that both the *sick* and the *well* populations are almost periodic in time in that as the number of well people initially increases, there will unfortunately be an accompanying increase in the number of sick people. The peak population of the sick people is delayed in time from the peak population of well people. There will be a time separation between the maximum total population which is also shown in Figure 2 and the maximum of either the well or sick populations will eventually decay to the equilibrium states given in (3).

Hopefully this cyclical nature of the populations of well and sick people will follow in the footsteps of a smallpox virus which was once the scourge of humanity eliminating the further need to study the smallpox sick population using this very simple model.

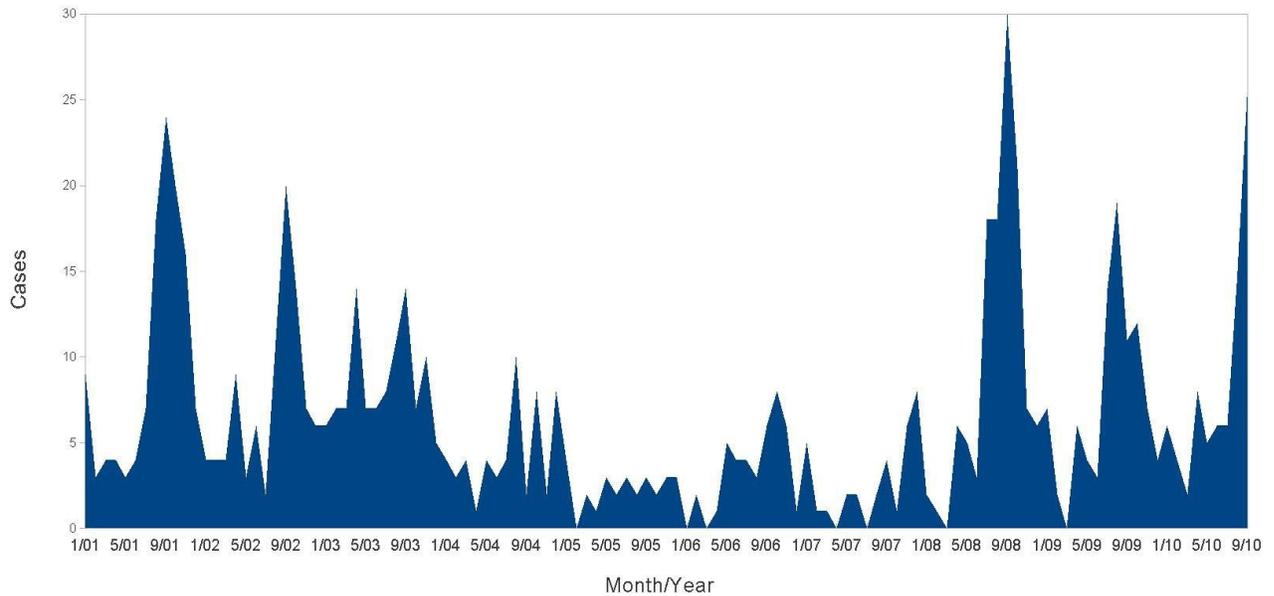

**Figure 1**.  Poliomyelitis case in Pakistan since 2001 until 2010.

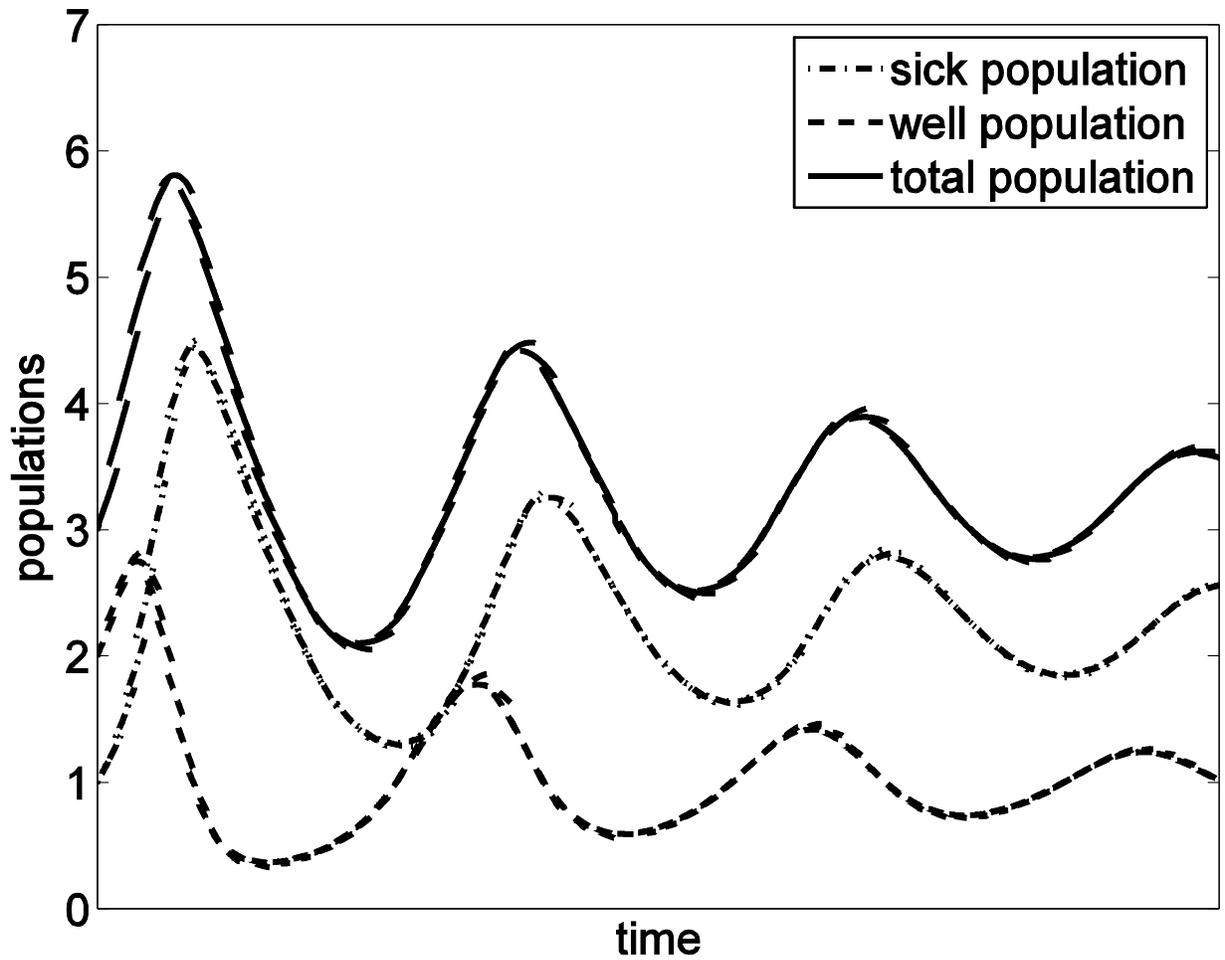

**Figure 2**. Simulation of the well population, the sick population and the total population as a function of time using equations (1) and (2) is depicted in the figure. The numerical values that were used to create this figure were arbitrarily chosen to be:

$\alpha$ = 0.2, $\beta$ = 0.1, $\gamma$ = 0.01, $\delta$ = 0.1, and $\varepsilon$ = 0.1. The initial populations were *well* = 2 and *sick* = 1 and the longtime asymptotic values are at *well* = 1 and *sick* = 2.22